\documentstyle[psfig,twocolumn,prl,aps,amssymb]{revtex}

\begin{document}
\draft

\date{\today} \title{The Phonon Drag Effect in Single-Walled Carbon 
Nanotubes} 
\author{V.W. Scarola and G.D. Mahan}  
\address{Department of Physics, 104 Davey Laboratory, The Pennsylvania State 
University, University Park, Pennsylvania 16802}
\maketitle
\begin{abstract}
A variational solution of the coupled electron-phonon Boltzmann 
equations is used to calculate the phonon drag contribution to the 
thermopower in a 1-D system.  A simple formula is derived for the temperature 
dependence of the phonon drag 
in metallic, single-walled carbon nanotubes.  
Scattering between different 
electronic bands yields nonzero values for the phonon drag as the 
Fermi level varies.  
\end{abstract}
\pacs{73.63.Fg,72.15.Jf}
\maketitle

\section{Introduction}

Since their discovery \cite{iijima}, carbon nanotubes 
have provided a testbed for fundamental and applied physics.  
Characterization of these new systems is crucial for 
future progress.   
The thermopower of a material is an intrinsic quantity 
which yields important information related to the 
electronic band structure, electron-phonon coupling parameters, 
and relaxation rates of the system.  
Recent experiments on mats  
of single walled carbon nanotubes 
\cite{mizelprl,romero,eklundkondo,eklundI,barisic,tian,choi}  
have found surprisingly large values for the 
thermoelectric power ($\sim 50 \mu V/ K$)
 under various ambient conditions.  When the thermopower is plotted 
versus temperature most
of the data show a quasi-linear behavior at temperatures 
greater than 200K.  Thermopower linear in temperature suggests that 
conduction through metallic tubes dominates the thermopower 
at these temperatures.  However, 
some data show nonlinearities below $\sim 200K$.  Here a peak 
in the thermopower is observed.  
The origin of this peak is the subject of much debate.  
Several effects that could lead to such behavior 
have been discussed in the literature.  These include 
parallel transport 
through semiconducting tubes  
\cite{mizelprl}, a one-dimensional Kondo 
effect \cite{eklundkondo}, 
and the phonon drag effect.  
Semiconducting tubes are not expected to contribute 
much to the thermopower because metallic tubes have 
a larger electrical conductivity \cite{romero}.
The Kondo effect has been invoked to explain the large 
peak in samples containing magnetic impurities.  However, a smaller, 
broad peak remains when the magnetic impurities are removed and 
therefore does not explain the universality of the low temperature 
nonlinearity in samples without magnetic impurities.    

Turning to the 
phonon drag effect, very little theoretical work has been done to 
calculate the phonon drag contribution to the thermopower in metallic 
carbon nanotubes.  Rough, low temperature, estimates often 
quote the result 
\begin{equation}
S_{\text{drag}}\propto C_{v}
\label{free}
\end{equation} 
where $C_v$ is the lattice specific heat.  However, this formula assumes 
that phonons scatter electrons within one 
parabolic band and that electron-phonon scattering acts as the 
dominant phonon decay mechanism.
Applying 
Eq.~(\ref{free}) to nanotubes with two bands, particle and hole,
 yields a negligible contribution 
to the thermopower due to drag because the particle and hole contributions 
cancel.   

The objective of this work is to extract the temperature 
dependence of the phonon drag contribution to 
the thermopower for a one dimensional system, within the  
linear band approximation.  We consider a model wherein
mechanisms other than electron-phonon scattering limit phonon 
lifetimes.  We derive a 
new expression for the phonon drag in these systems using 
a solution of the coupled electron-phonon Boltzmann 
equations.   We find that {\em inter}band 
scattering gives a non-zero contribution to the thermopower when the 
Fermi level does {\em not} lie at the band crossing.  Interband 
transitions near the Fermi level contribute appreciably to the 
phonon drag thermopower.
Our results can be summarized in the following formula
\begin{eqnarray}
S_{\text{total}}=AT+\frac{B}{T^{m+1}}
\frac{\text{sign}(-\mu)}{e^{\frac{2c|\mu|}{vk_BT}}-1}
\left(1+\frac{c|\mu|}{vk_BT}\right)
\label{total}
\end{eqnarray}
in the limit
\begin{eqnarray}
k_BT\ll|\mu|\ll\frac{k_BT_Dv}{2c}
\nonumber
\end{eqnarray} 
where $A$ and $B$ are fitting parameters, 
$\mu$ is the energy difference between 
the Fermi level and the band crossing, $v$ the electron speed,  
$c$ a typical phonon speed, and $T_D$ the Debye temperature.  $m$ 
parameterizes the temperature dependence of the phonon relaxation 
time.  In metals one typically finds $m\sim1$ at temperatures near the Debye 
temperature.  
The term linear in temperature 
is the usual diffusive contribution for metals and the second term is our 
result for the phonon drag part at low temperatures.  
This formula yields a peak, due to phonon drag.   
We therefore associate the phonon drag effect with 
the low temperature nonlinearities observed in thermopower measurements 
on single walled carbon nanotubes.

The plan of the paper is as follows.  In Section II we outline the 
basic theory of thermopower measurements in metals.
In Section III Bailyn's theory of phonon drag in 
metals is reviewed.  In Section IV we apply Bailyn's theory to a 
simple 1-D model of single walled, (10,10) armchair carbon nanotubes.  
Here we derive an
expression for the phonon drag in these systems.  In Section V we
 extract a simple formula by 
making a low temperature approximation.  
In Section VI we conclude by discussing the limits of our approximations.

\section{Thermopower Due To Electron Diffusion and Phonon Drag}

The application of a temperature gradient to a metal leads to the diffusion
of charge carriers from the warm to the cold end of the sample.  
The thermopower measures the charge build up across the sample. 
It is given by $S=\lim_{\Delta T \rightarrow 0} \Delta V /\Delta T$, where 
$\Delta V$ and $\Delta T$ are the potential and temperature differences 
across the sample, respectively. 
The thermopower due to diffusion, $S_{\text{diff}}$, 
can be calculated from the standard 
set of transport coefficients.    
The transport coefficients can in turn be derived from the Boltzmann equation 
for the electron distribution function.  A general argument leads 
to the Mott expression for the thermopower due to diffusion
\begin{eqnarray}
S_{\text{diff}}=\frac{-\pi^2}{3}\frac{k_B^2T}{|e|}
\left( \frac{\partial \text{ln} \sigma}{\partial E} \right )_{\text{EF}} 
\end{eqnarray}
where $E$ is the energy, $\sigma$ the electrical conductivity, and 
$EF$ the Fermi energy.  When there is more than one band present the 
thermopower due to each band adds
\begin{eqnarray}
S_{\text{diff}}=\frac{\sum_{\ell}\sigma_{\ell}S_{\text{diff}}^{\ell}}
{\sum_{\ell}\sigma_{\ell}}
\end{eqnarray}
where $\ell$ is the band index.  

We apply the above equations to the case of metallic carbon nanotubes 
by considering electrons in one dimension.  The electronic states 
fill two overlapping, parabolic bands, the particle and hole bands, 
up to the Fermi level.  
With only one parabolic band we expect the diffusive part of the
thermopower in a metal to vary linearly with temperature.  The 
factor $\left( \frac{\partial \text{ln} \sigma}{\partial E} \right )_{EF} $  
depends on the details of the system, including the 
density of states.  However, with two bands present, 
the thermopower due to 
states filling the hole band 
cancels the contribution from the particle band when the 
Fermi level lies at the band crossing.  Within this 
approximation we have no net contribution to the thermopower 
in a metallic tube. 

When the Fermi level is allowed to move within the rigid $\pi$ bands 
the resulting thermopower is non-zero \cite{romero}.
Recent calculations show that an enhancement 
in the density of states due to impurities \cite{romero,lammert,kostryko} 
or tube-tube interactions \cite{mizelprl,cohen2,kwon} may generate 
large contributions to the thermopower through the term 
$\left( \frac{\partial \text{ln} \sigma}{\partial E} \right )_{EF} $.  
We therefore consider the following 
standard form for the diffusive contribution to the thermopower in metals
\begin{eqnarray}
S_{\text{diff}}=AT
\end{eqnarray}
where the constant $A$ is a fitting parameter 
which may vary with the Fermi level.  Recent experiments 
on mats of single walled, carbon nanotube bundles do indeed show this 
behavior at large temperatures ($T > 200K$) \cite{romero}.  

An anomalous peak in the thermopower appears at low temperatures 
in several different experiments on single-walled and multi-walled 
carbon nanotubes.
In standard thermopower measurements of 3-D metals such behavior is 
often associated with the phonon drag effect whereby the phonon flux from 
the hot end of the sample to the cold end drags additional charge 
carriers to the cold end of the sample via momentum transfer.  
This effect adds to the thermopower in conventional
metals  
\begin{equation}
S_{\text{total}}=S_{\text{diff}}+S_{\text{drag}}
\end{equation}
Standard, low temperature estimates of the phonon drag 
contribution to the thermopower rely on the relation
\begin{equation}
S_{\text{drag}}\simeq\frac{-C_vt}{3n|e|}
\label{drag0}
\end{equation}
where $C_v$ is the lattice specific heat, $n$ the carrier density, and 
the $t$ is the transfer factor.   $t$ is a rough estimate 
of the probability that a phonon collides with an electron relative to 
all scattering events.  In one dimension the lattice specific 
heat is nearly linear in temperature 
\cite{mizelprb,cohen} and does not provide the nonlinear
 temperature dependence 
required to explain the peak observed in measurements on carbon nanotubes 
\cite{caveat}.  
Moreover, if we consider only intraband scattering, 
the contributions from states filling the electron and 
hole bands should cancel to give no net drag.  

The above formula for the drag contribution cannot be applied to metallic  
carbon nanotubes for two reasons.  The derivation of the above formula 
\cite{blatt} relies on a free, electronic band structure where transitions 
lie only within the parabolic bands.  
It also assumes that 
the dominant decay mechanism for phonons is electron-phonon scattering.  
Below we derive a new formula for the low temperature phonon 
drag contribution to the thermopower in one dimension.  
We find a nonzero contribution to the 
phonon drag part of the thermopower when we include transitions between 
two linear bands and
assume that the dominant decay mechanism for phonons is {\em not} 
electron-phonon scattering.

\section{Bailyn Formalism for Phonon Drag}

The phonon drag contribution to the thermopower in a metal 
may be calculated by solving the coupled electron-phonon Boltzmann 
equations.  In this section we briefly review Bailyn's 
formalism for calculating 
the phonon drag.  Following Ref.~\cite {Bailyn58} we write the Boltzmann 
equation for the electron distribution function, $f$, in the 
relaxation time approximation.  Using first oder perturbation theory for 
the electron transition probabilities one finds

\begin{eqnarray}
&(&\frac{\partial f}{\partial t})_{\text{coll}}
=\sum_{\vec{k}', \ell'}C_{\vec{k},\vec{k}'j}
\{\delta(-)\delta_{\vec{k}',\vec{k}+\vec{q}}
\nonumber
\\
&[&-N(\vec{q}j)f(1-f')+(N(\vec{q}j)+1)f'(1-f)]
\nonumber
\\
&+&\delta(+)\delta_{\vec{k}',\vec{k}+\vec{q}}[-(N(\vec{q}j)+1)f(1-f')
\nonumber
\\
&+&N(\vec{q}j)f'(1-f)]\}
\nonumber
\\
&-&(f-f_0)/\tau(\vec{k})
\end{eqnarray}
where $N(\vec{q}j)$ is the phonon distribution function,
 $\omega=c|q|$ is the frequency of a phonon with speed $c$, 
$f_0$ is the
Fermi-Dirac distribution function,  and $\tau(\vec{k})$ is the relaxation
time due to the electron-electron interaction. 
We define the energy delta functions
\begin{eqnarray}
\delta(\pm)\equiv\delta(E(\vec{k}'\ell')-E(\vec{k}\ell)\pm \hbar\omega)
\end{eqnarray}
The factor $C_{\vec{k},\vec{k}'j}$ is related to the electron-phonon matrix 
elements  
\begin{eqnarray}
C_{\vec{k},\vec{k}'j}=\frac{{\cal M}(\vec{q}j)}{2\hbar\omega(\vec{q}j)}
\end{eqnarray}
where ${\cal M}(\vec{q}j)\equiv 
|<\vec{k}'\ell'|\vec{\nabla} U\cdot\vec{\epsilon}(\vec{q}j)|\vec{k}\ell>|^2
/MN_c$ is the square of the matrix element for the scattering of an electron 
from wave vector $\vec{k}$ and band $\ell$ to 
wave vector $\vec{k}'$ and band $\ell'$
by a phonon of wave vector $\vec{q}$
and polarization $j$.  
Here $\vec{\nabla} U$ is the gradient of the ion potential, $N_c$ 
the number of cells in the periodic block, $M$ the ion Mass, and 
$\vec{\epsilon} (\vec{q}j)$ the phonon polarization
vector.
The above matrix element ignores Umklapp scattering \cite{caveat2}.

From the above expression for the electron distribution 
function we can read off the necessary terms for the phonon Boltzmann 
equation.  The second and third terms show gains in the phonon 
distribution.  The first and fourth terms show losses in the phonon 
distribution function.  This gives 
 \begin{eqnarray}
&-&(\frac{\partial N(\vec{q}j)}{\partial t})_{\text{drift}}=
(\frac{\partial N(\vec{q}j)}{\partial t})_{\text{coll}}
\nonumber
\\
&=& \sum_{\vec{k}, \vec{k}'}C_{\vec{k},\vec{k}'j}
\{\delta(-)\delta_{\vec{k}',\vec{k}+\vec{q}}[-N(\vec{q}j)f(1-f')
\nonumber
\\
&+&(N(\vec{q}j)+1)f'(1-f)]-\delta(+)
\delta_{\vec{k}',\vec{k}+\vec{q}}
\nonumber
\\
&[&-(N(\vec{q}j)+1)f(1-f')+N(\vec{q}j)f'(1-f)]\}
\nonumber
\\
&-&(N(\vec{q}j)-N_0)/\tau(\vec{q})
\end{eqnarray}
where $\tau(\vec{q})$ is the phonon relaxation time and $N_0$ is the Bose 
distribution.
The above two equations for the electron and phonon distribution functions 
can be solved using a variational procedure.  From 
the relevant transport coefficients the phonon drag contribution to the 
thermopower can be extracted.  The most general form for which was 
derived in Ref.~\cite{Bailyn67}.  
\begin{eqnarray}
S_{\text{drag}}&=&\frac{2|e|k_B}{\sigma d}\sum_{\vec{q}j} 
\frac{\partial N_0(\vec{q}j)}{\partial T}
\nonumber
\\
&\times&\sum_{\vec{k}\ell; \vec{k}'\ell'} \alpha(\vec{q}j;\vec{k}\ell, \vec{k}'\ell') 
[\vec{v}_{\vec{k}\ell}\tau_{\vec{k}\ell}-\vec{v}_{\vec{k}'\ell'}\tau_{\vec{k}'\ell'}]\cdot \vec{V}_{\vec{q}j}
\label{drag}
\end{eqnarray}
Here $\sigma$ is the electrical conductivity, $d$ the dimensionality of 
the system, the 2 results from a sum over the spin degrees of freedom,
$\vec{v}_{\vec{k}\ell}$ is the electron group velocity, and 
$\vec{V}_{\vec{q}j}$ is the phonon 
group velocity.  The factor 
$\alpha$ is the relative probability that the $\vec{q}j$ phonon will 
scatter an electron from the state $\vec{k}\ell$ to the state $\vec{k}'\ell'$, relative 
to all other possible phonon collisions.

Details of the electron-phonon interaction are included in $\alpha$.  
Symbolically
\begin{eqnarray}    
\alpha=\frac{\tau_{ep}^{-1}}{\sum\tau_{ep}^{-1}+\tau_{p}^{-1}}        
\end{eqnarray}
where $\tau_{ep}$ is the phonon relaxation time due to the electron-phonon 
interaction and $\tau_{p}$ is the phonon relaxation time due to 
any other interaction.  This may include phonon-phonon, phonon-boundary, 
phonon-impurity, or phonon-defect scattering.  Using the above results for the 
phonon relaxation rates Bailyn finds \cite{Bailyn60}
\begin{equation}
\alpha(\vec{q}j;\vec{k}\ell, \vec{k}'\ell')
=\frac{I_{\vec{k}\ell,\vec{k}'\ell'}}
{\frac{T}{\hbar\omega\tau_p(\vec{q})}\frac{\partial N_0}{\partial T}
+\sum_{\vec{k}\ell,\vec{k}'\ell'}I_{\vec{k}\ell,\vec{k}'\ell'}}
\label{alpha}
\end{equation}
where
\begin{eqnarray}
I_{\vec{k}\ell,\vec{k}'\ell'}&=&\frac{1}{2\hbar\omega k_B T} f_0(E(\vec{k}\ell))\{1-f_0(E(\vec{k}'\ell'))\}
\nonumber
\\
&N&_0(\vec{q}j){\cal M}(\vec{q}j)\delta(-)\delta_{\vec{k}',\vec{k}+\vec{q}}
\label{I}
\end{eqnarray}
Equations (\ref{drag}),(\ref{alpha}),and (\ref{I}) constitute Bailyn's 
theory of phonon drag 
in metals.  The remainder of this article will be concerned 
with the application of this formalism to the case of metallic carbon 
nanotubes. 

\section{Bailyn Formalism Applied to Metallic Carbon Nanotubes}

We consider a one dimensional lattice lying on the 
$z$ axis with left and right moving electrons with the 
band structure of 
a (10,10) armchair carbon nanotube.  
It is assumed that 
the electron relaxation time has a weak wave vector dependence 
for transitions about the K point ($2\pi/3a$ for lattice spacing $a$) 
so that
$\tau(\vec{k}\ell+2\pi/3a)\sim\tau(2\pi/3a)$.  
  Our expression 
for the phonon drag part of the thermopower then reads
\begin{eqnarray}
S_{\text{drag}}&=&\frac{2|e|k_B\tau}{\sigma}\sum_{qj} 
\frac{\partial N_0(qj)}{\partial T}
\nonumber
\\
&\times&\sum_{\vec{k}\ell; \vec{k}'\ell'} \alpha(q,j;\vec{k}\ell, \vec{k}'\ell') 
[\vec{v}_{\vec{k}\ell}-\vec{v}_{\vec{k}'\ell'}]\cdot \vec{V}_{qj}
\label{drag4}
\end{eqnarray}
We note in passing that ignoring all but electron-phonon scattering 
reduces the drag to  
Eq.~(\ref{free}) when we take $\alpha=\delta_{\vec{k}',\vec{k}+\vec{q}}$ 
\cite{blatt}.  
The resulting sum over 
$k,k'$ becomes $\hbar\omega$.  We assert that this limit is not applicable 
to metallic carbon nanotubes.

\begin{figure}
\centerline{\psfig{figure=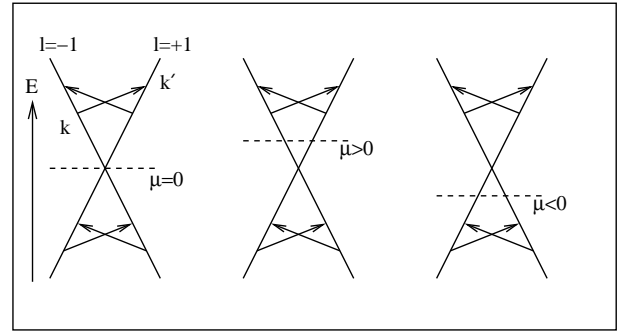,width=3.5in,angle=-90}}
\caption{The linear band model depicting energy along the 
y-axis and wave vector along the x-axis.  The two bands shown, 
$\ell=+1$ and $-1$, cross at zero energy.  The arrows indicate possible 
interband transitions from an initial electron state $k$ to a final
state $k'$.  The dashed line indicates the position of the 
Fermi energy relative to the band crossing point, defined to be $\mu$.
}
\label{bands}
\end{figure}

To make progress we evaluate Eq.~(\ref{drag4}) using approximations valid 
for metallic carbon nanotubes.  In these systems 
it is reasonable to assume that at small wave vectors and energies 
 only acoustic 
phonons scatter electrons and that their dispersion is linear 
\cite{eklundreview}.  We then have
\begin{eqnarray}
 \vec{V}_{qj}=c_j \text{sign}(q) \hat{z}
\end{eqnarray}
where $c_j$ is the phonon speed in the $j$th band.
The electronic band structure consists of two nearly linear 
bands crossing at zero energy, Fig.~\ref{bands}.  
This approximation holds as 
long as the Fermi level lies no more than $\sim1eV$ from the 
band crossing.  
Above $1eV$ other bands will complicate the spectrum.  For carrier 
excitations around the K point  we have 
\begin{equation}
E(k\ell)=\ell\hbar vk
\end{equation}
where $v$ is the Fermi speed and $\ell=+1,-1$ labels the two 
bands.
With the linear band approximation the electron group velocity is then
\begin{eqnarray}
\vec{v}_{k\ell}= \frac{1}{\hbar}\frac{\partial E(k\ell)}{\partial k}\hat{z}
=v \ell \hat{z}
\end{eqnarray}
Substituting the electron and phonon velocities into the expression 
for $S_{\text{drag}}$ gives
\begin{eqnarray}
S_{\text{drag}}&=&\frac{2|e|k_B\tau v}{\sigma}\sum_{qj} c_j 
\text{sign}(q) \frac{\partial N_0(qj)}{\partial T}
\nonumber
\\
&\times&\sum_{k\ell; k'\ell'} \alpha(qj;k\ell, k'\ell') 
[\ell-\ell']
\end{eqnarray}
Note that the above expression for the phonon drag vanishes 
when only $intra$band scattering is allowed, i.e. $\ell=\ell'$.  
In what follows we consider only $inter$band scattering, $\ell\neq\ell'$.  
Define the transfer factor to be
\begin{eqnarray}
t(q)\equiv\sum_{k\ell; k'\ell'} \alpha(qj;k\ell, k'\ell') 
[\ell-\ell']
\end{eqnarray}
where $\alpha$ is given by Eqs.~(\ref{alpha}) and (\ref{I}).  
The drag then has the simple form
\begin{eqnarray}
S_{\text{drag}}=\frac{2|e|k_B\tau v}{\sigma}\sum_{qj} c_j 
\text{sign}(q) \frac{\partial N_0(qj)}{\partial T} t(q)
\label{drag2}
\end{eqnarray}

To find $t(q)$ we need to evaluate the following expression
\begin{eqnarray}
\sum_{k\ell,k'\ell'}\alpha[\ell-\ell']= 
2\frac{\sum_{kk'}I_{k,1;k',-1}-\sum_{kk'}I_{k,-1;k',1}}
{\frac{T}{\hbar\omega\tau_p(q)}\frac{\partial N_0}{\partial T}
+\sum_{\vec{k}\ell,\vec{k}'\ell'}I_{\vec{k}\ell,\vec{k}'\ell'}}
\end{eqnarray}

To simplify the expression for $t(q)$ we make an assumption regarding  
the available phonon scattering processes. 
Note that phonons traveling along a single nanotube may scatter through 
many different mechanisms.  The large amount of surface area exposes the 
phonons to impurities, defects, and neighboring tubes.  
Furthermore, small tube lengths, $\sim 10 \mu m$, allow for 
phonon-boundary scattering.
We consider here two simple forms for the phonon 
relaxation time, valid at low and high temperatures, respectively.
At low temperatures, the phonon decay mechanism is 
a competition between boundary and defect scattering which add
no temperature dependence to the phonon relaxation time.  
At high temperatures, near room temperature, one expects 
phonon-phonon scattering to contribute significantly to phonon decay.  
Anharmonic scattering will require three phonons.  The 
third being an optical mode at long wavelengths.  The 
scattering rate will be proportional to the 
number of available optical phonons.  One can then show that 
the relaxation time is inversely proportional to temperature, at 
large temperatures.
We therefore take the following form for the phonon relaxation time 
\begin{eqnarray}
\tau_p = \tau_0 \left ( \frac{\phi}{T} \right )^m
\label{relax}
\end{eqnarray}
where $\phi$ is a characteristic temperature, $m=0$ for boundary/defect
scattering, 
and $m=1$ for phonon-phonon scattering.  Comparison with experiment 
will require a more accurate expression for $\tau_p$.

We now invoke the assumption that mechanisms other than electron-phonon
scattering limit the phonon lifetime, 
i.e. $\tau_p$ is small \cite{limit}. 
More precisely 
\begin{eqnarray}
\frac{T}{\hbar\omega\tau_p(q)}\frac{\partial N_0}{\partial T}
\gg\sum_{k\ell,k'\ell'}I_{k\ell,k'\ell'}
\end{eqnarray}
The above approximation enters phonon drag studies of quantum 
wires in GaAs.  Analogous results for the phonon drag were 
also obtained in Ref.~\cite{butcher1}.    
The transfer factor becomes
\begin{eqnarray}
t(q)&\simeq&\frac{2\hbar\omega\tau_p(q)}{T\frac{\partial N_0}{\partial T}}
\nonumber
\\
&\times&
\left [\sum_{k,1;k',-1}I_{k,1;k',-1}-\sum_{k,-1;k',1}I_{k,-1;k',1}\right ] 
\end{eqnarray}
To further evaluate the transfer factor we return to the matrix element in 
Eq.~(\ref{I}).  Working with the deformable ion model at low wave vectors
($q\ll q_{\text{Debye}}$) one finds \cite{Bailyn67}
\begin{equation}
{\cal M}(qj)=D_j (\hbar\omega_j)^2
\label{matrix}
\end{equation}
where $D_j$ is a constant depending on the ion mass, 
the deformation energy, 
and other tube parameters including the radius and lattice spacing.  This 
constant has been evaluated for (10,10) carbon nanotubes \cite{mahan}.  
The low wave vector 
approximation made above has been motivated by measurements of the Debye 
temperature.  It has been measured to be near 1000K in 
metallic carbon nanotubes \cite{hone}.  It will be shown that, at 
low temperatures,  the 
temperature dependence of the drag does not depend on the 
precise form of the matrix elements ${\cal M}$.      

The transfer factor contains two terms of the form 
\begin{eqnarray}
&\sum&_{k,k'}I_{k\ell,k'\ell'}=\frac{{\cal M}(qj)}{2 k_B T \hbar \omega } 
\nonumber
\\
&\times&\sum_{k,k'}f_0(E(k\ell))
\{1-f_0(E(k'\ell'))\}N_0(qj)
\nonumber
\\
&\times&\delta(-)\delta_{k',k+q}
\nonumber
\\
&=&\frac{{\cal M}(qj)}{2 k_B T \hbar \omega} 
\nonumber
\\
&\times&\sum_{k}f_0(E(k\ell))
\{1-f_0(E(k+q,\ell'))\}N_0(qj)
\nonumber
\\
&\times&\delta(E(k+q,\ell')-E(k\ell)-\hbar\omega)
\label{sum}
\end{eqnarray}
Where the Fermi-Dirac distribution function is  
$f_0(E)=[e^{(E-\mu)/k_BT}+1]^{-1}$.
The energy delta function in  Eq.~(\ref{sum}) needs to be evaluated 
for two cases.  
The first case is $\ell=1$ and $\ell'=-1$. We then have
\begin{eqnarray}
&E&(k+q,-1)-E(k,1)-\hbar\omega
\nonumber
\\
&=&\hbar[-v(k+q)-vk-c|q|]=0
\end{eqnarray}
This has a solution $k^a\equiv -\frac{|q|}{2}[\text{sign}(q)+\frac{c}{v}]$.
The second case is $\ell=-1$ and $\ell'=1$.  The solution for 
this case is $k^b\equiv -\frac{|q|}{2}[\text{sign}(q)-\frac{c}{v}]$.
Upon a change of variables in the energy delta function of 
Eq.~(\ref{sum}) the sum can be evaluated.  The transfer factor 
now becomes 
\begin{eqnarray}
t(q)&=&\frac{\tau_p(q)}{\frac{\partial N_0}{\partial T}}
\left(\frac{{\cal M}(qj)}{k_B T^2}\right ) 
\left(\frac{N_0(qj) L}{4\pi\hbar v}  \right )
\nonumber
\\
&\times&[f_0(\hbar v k^a)\{1-f_0(\hbar v k^b)\}
\nonumber
\\
&-&f_0(-\hbar v k^b)\{1-f_0(-\hbar v k^a)\}] 
\end{eqnarray}  
where 
$L$ is the tube length.
Substituting the transfer factor into our expression for the drag, 
Eq.~(\ref{drag2}), gives
\begin{eqnarray}
S_{\text{drag}}&=&\frac{|e|\tau v}{\sigma T^2}\frac{L}{2\pi k_B \hbar v}
\sum_{qj} c_j \text{sign}(q) {\cal M}(qj) \tau_p(q) N_0(qj)
\nonumber
\\
&\times& [f_0(\hbar v k^a)\{1-f_0(\hbar v k^b)\}
\nonumber
\\
&-&f_0(-\hbar v k^b)\{1-f_0(-\hbar v k^a)\}]  
\end{eqnarray}
To evaluate the sum over $j$ it is convenient to 
assume that only one linear phonon branch contributes to the thermopower.  
The inclusion of other linear modes with approximately the same 
phonon speed will simply add to the overall constant.  Passing to 
the continuum limit and imposing a Debye cutoff, $q_D$, one can show that
 \begin{eqnarray}
S_{\text{drag}}&=&\frac{|e|\tau c L^2}{2 \pi^2 \sigma T^2 k_B \hbar}
\int_{0}^{q_D}dq {\cal M}(q) \tau_p(q)N_0(q)
\nonumber
\\
&\times& [f_0(\hbar v k^a)\{1-f_0(\hbar v k^b)\}
\nonumber
\\
&-&f_0(-\hbar v k^b)\{1-f_0(-\hbar v k^a)\}]  
\end{eqnarray} 
To further simplify the drag formula we make a change of variables with the 
following definitions
\begin{eqnarray}
u&\equiv& \frac{\hbar cq}{2k_B T}\left(\frac{v}{c}-1\right)
\nonumber
\\
\overline f(x)&\equiv&\frac{1}{e^{(x-\mu/k_BT)}-1}
\nonumber
\\
\gamma&\equiv&\frac{v+c}{v-c}
\end{eqnarray}
The drag can then be written as
\begin{eqnarray}
S_{\text{drag}}&\simeq&\frac{|e|\tau c L^2}{\pi^2 \sigma \hbar^2 T v}
\int_{0}^{\frac{T_Dv}{2cT}} du 
\frac{{\cal M}\tau_p }{e^{2uc/v}-1}  
\nonumber
\\
&\times&[\overline 
f(-u\gamma)\{ 1- \overline f(-u)\}
-\overline f(u)\{ 1- \overline f(u\gamma)\} ]
\label{drag3}
\end{eqnarray}
where the Debye temperature is defined as usual $T_D\equiv\hbar c q_D/k_B$.  
In the above we have used the fact that 
for (10,10) arm chair carbon nanotubes 
$c=20.35\times10^3$ m/s \cite{eklundreview}
for the longitudinal acoustic mode whereas $v=8.4\times10^5$ m/s.  This 
gives $\frac{v}{c}\sim 100$. 
Equation (\ref{drag3}) is our primary result and contains several interesting 
features.  First note that when $\mu=0$ we have $\overline f(-x) 
= 1-\overline f(x)$, in which case the two terms in the integral cancel 
giving no contribution to the thermopower.  If we take $\mu >0$ 
then the second term dominates, giving a negative contribution to the 
thermopower.  Similarly, if we take $\mu<0$ we get a positive contribution 
to the thermopower.  

The overall sign of $S_{\text{drag}}$ can be deduced from Fig.~\ref{bands}.  
For $\mu>0$, interband transitions above the crossing point are favored. 
Transitions with $q>0$ yield a positive change in 
the electron group velocity which, from Eq.~(\ref{drag4}), 
give an overall negative sign to the drag.  
For $\mu<0$, interband transitions below the crossing point are favored 
which give and overall positive sign to the drag.      

The Debye temperature plays a small role because 
the kernel of the integral is sharply peaked.  As long as the 
peak, $u_{pk}=\frac{\mu}{k_BT}$, lies within the range of integration the Debye temperature can be set 
to infinity.  When the peak crosses the range of integration, 
$\mu \sim \frac{k_B T_D v}{2c}$,  
$S_{\text{drag}}$ falls to zero.  For, $\mu >  \frac{k_B T_D v}{2c}$
 the set of transitions favored by the electron distribution 
require wave vectors above the Debye cutoff, $q_D$.

\begin{figure}
\centerline{\psfig{figure=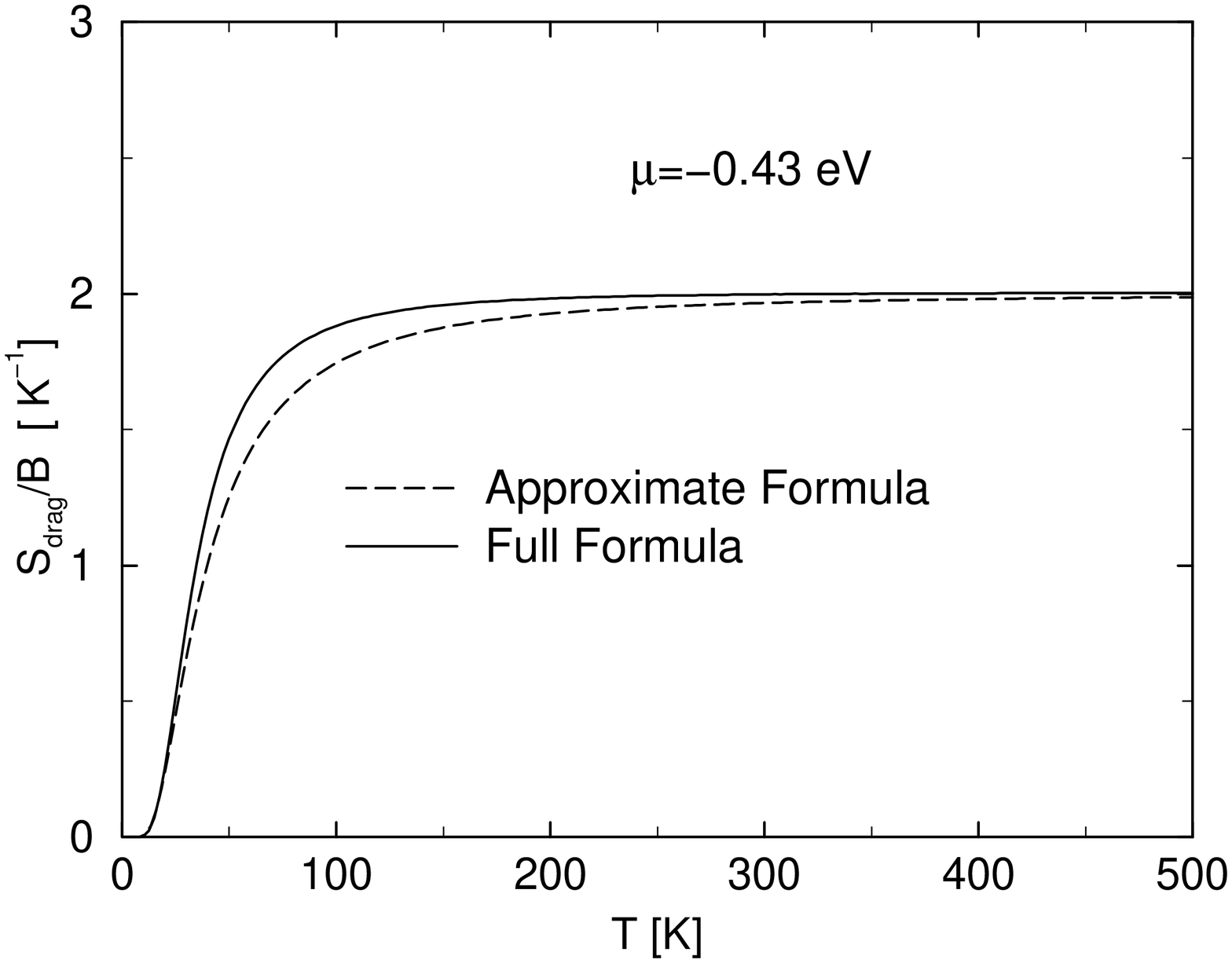,width=4.0in,angle=-90}}
\caption{  
The phonon drag contribution to the thermopower plotted, 
in units of the constant $B$, 
as a function of temperature.  The 
Fermi energy is chosen to be  
$\mu=-0.43$eV.
The ratio between the electron and phonon speeds is 
taken to be $\frac{v}{c}=100$.  The solid line shows the 
result from the full formula, Eq.~(\ref{drag3}) with $T_D=1000K$.  
The dashed line 
shows the simplified approximation, the second term in 
Eq.~(\ref{total}). 
For purposes of comparison, a temperature independent phonon 
relaxation time is assumed, 
$m=0$ in Eq.~(\ref{relax}). 
}
\label{approx}
\end{figure}

Fig.~\ref{approx} shows the temperature dependence of $S_{\text{drag}}$. 
To evaluate the 
integral we have taken the deformable ion model, Eq.~(\ref{matrix}).
We have also taken a temperature independent phonon relaxation time
, $m=0$ in Eq.~(\ref{relax}).  
For large 
temperatures $S_{\text{drag}}$ flattens because 
we assume here that $\tau_p$ does not depend on temperature.  
In real samples it is likely that, for large 
temperatures, phonon-phonon scattering 
may impose a temperature dependence 
to the phonon relaxation time.  We would then see a change in the 
drag at large temperatures.  In Fig.~\ref{compare} we compare 
the temperature dependence of the drag for the two cases $m=0$ and 
$m=1$ in Eq.~(\ref{relax}).  Here we have taken $\phi=100 K$, $v/c=100$, 
$\mu=0.436$ eV, and $T_D=1000 K$.  A suppression of the drag 
at large temperatures clearly induces a peak, similar to those observed in 
experiments.
The low temperature behavior can be extracted with a few approximations.  

\section{The Low Temperature Limit}

A low temperature approximation to the integral formula for $S_{\text{drag}}$, 
Eq.~(\ref{drag3}), can be obtained for nonzero values of $\mu$.  For $\mu<0$ 
the second term in the integrand vanishes.  We then have
\begin{eqnarray}
&S&_{\text{drag}}\simeq\frac{ |e|\tau c L^2}{\pi^2 \sigma \hbar^2 T v}
\nonumber
\\ 
&\times&\int_{0}^{\frac{T_Dv}{2cT}} du 
\frac{{\cal M}\tau_p }{e^{2uc/v}-1} \left [\overline 
f(-u\gamma)\{ 1- \overline f(-u)\}\right]
\end{eqnarray}
As mentioned earlier the factor $\overline f(1-\overline f)$ in the 
 integral is a sharply peaked function centered at the value 
$u_{pk}\approx\frac{\mu}{k_BT}$.  Thus 
\begin{eqnarray}
&S&_{\text{drag}}\simeq
\frac{|e|\tau c L^2}{\pi^2 \sigma \hbar^2 T v}
\frac{{\cal M}(2\mu/\hbar v) \tau_p(2\mu/\hbar v)}{e^{2u_{pk}c/v}-1}
\nonumber
\\
&\times& \int_{0}^{\frac{T_Dv}{2cT}} 
du \left [\overline 
f(u)\{ 1- \overline f(u\gamma)\} \right ]
\end{eqnarray} 
The integral can now be performed giving
\begin{eqnarray}
 &\int&_{0}^{\frac{T_Dv}{2cT}} 
du \left [\overline 
f(u)\{ 1- \overline f(u\gamma)\} \right ]
\nonumber
\\
&=& 1+\frac{\mu c}{vk_BT}+\vartheta(\{\gamma-1\}^2)
\end{eqnarray} 
The above argument 
remains the same for the case $\mu>0$ except for an overall sign change.  
The low temperature drag contribution to the thermopower becomes 
the second term in Eq.~(\ref{total}),
where the temperature independent factor
$B\equiv \frac{|e|\tau c L^2}{\pi^2 \sigma \hbar^2 v} {\cal M}(2\mu/\hbar v)
\tau_0(2\mu/\hbar v)\phi ^m$
may be taken as a fitting parameter.    
Note that the low temperature limit 
is easily reached because $\frac{v}{c}\gg1$.  Fig.~\ref{approx}
compares the above approximation to the full formula 
given by Eq.~(\ref{drag3}).

Fermi statistics restricts the interband transitions to lie near the 
Fermi level, thereby excluding all but a narrow range 
of phonon wave vectors.  
Eq.~(\ref{total}) therefore applies to systems with any, 
non-singular $q$ dependence in the electron-phonon 
matrix elements.  Only 
the constant $B$ will change with different 
forms of electron-phonon coupling.    

Fitting Eq.~(\ref{total}) to measurements of $S_{\text{total}}$
versus temperature yields a rough estimate 
of $\mu$.  
Fig.~\ref{total_plot} shows the total thermopower for 
three possible values of $\mu$.  
Motivated by experiments on mats of single-walled 
carbon nanotubes \cite{romero} we assume that the 
parameter $A$ changes sign with the Fermi level.  We 
have kept the ratio $|\frac{A}{B}|$ fixed.  In experiments 
we expect the parameters $A$ and $B$ to have a nontrivial 
dependence on $\mu$ due to sample dependent variations 
in the phonon and electronic density of states.

\section{Conclusion}

We have shown that within a one dimensional model phonon drag resulting 
from interband transitions between two linear 
bands gives a nonlinear temperature dependence to the thermopower 
when the Fermi level does not lie at the band crossing.  
Assuming that mechanisms other than the electron-phonon interaction 
contribute to phonon scattering we derive 
a simple expression for the phonon drag contribution to the 
thermoelectric power in a model that approximates parameters found in 
single walled (10,10) carbon nanotubes.  The strength of the effect depends 
strongly on the position of the Fermi level.  

The above results suggest that the phonon drag effect is a good 
candidate for recent, low temperature anomalies in thermopower 
measurements on single walled carbon nanotubes.  We note, however, 
that, as is typical in theories of the phonon drag effect 
in conventional metals, the size of the fitting parameter $B$ is 
difficult to estimate from first principles.  

\begin{figure}
\centerline{\psfig{figure=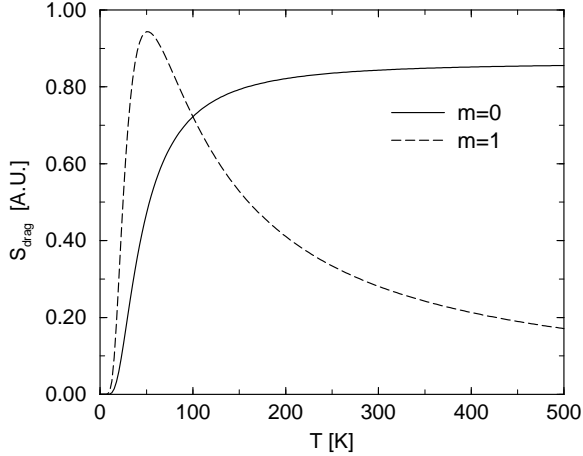,width=4in,angle=-90}}
\caption{  
The phonon drag contribution to the thermopower plotted versus 
temperature for two different cases.  The solid line shows 
the drag when a temperature independent phonon scattering mechanism 
dominates phonon decay, $m=0$ in Eq.~(\ref{relax}).  
The dashed line shows the 
drag when phonon-phonon scattering dominates, 
$m=1$.  The Fermi level is chosen to be 
$\mu=-0.5$ eV.  The parameter $B$ is the same for both curves.  
We also take $\frac{v}{c}=100$ and $\phi=100K$.
}
\label{compare}
\end{figure}

\begin{figure}
\centerline{\psfig{figure=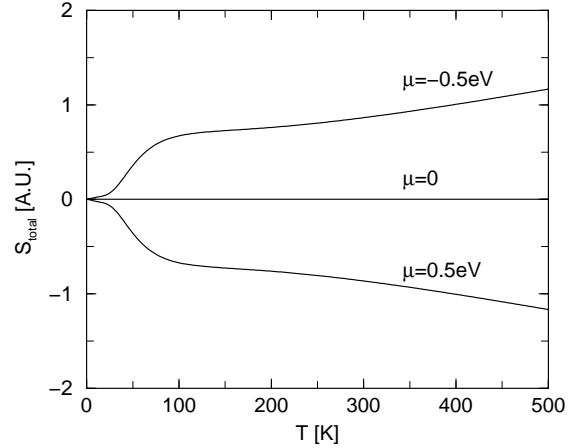,width=4.0in,angle=-90}}
\caption{  
The total thermopower, Eq.~(\ref{total}), plotted versus temperature 
for two values of the Fermi level, $\mu=-0.5$eV and $0.5$eV.  
The ratio of the two fitting parameters is taken to be 
$|A/B|=10^5$.  The top curve has a positive value for $A$ while 
the bottom curve has a negative value for $A$.  
We also have $\frac{v}{c}=50$, $\phi=100 K$, and 
$m=1$.  The central line is the $\mu=0$ case in Eq.~(\ref{drag3}), 
where $A=0$.   For $\mu=0$ transitions above and below the 
Fermi level cancel to give no net thermopower.   
}
\label{total_plot}
\end{figure}

In applying the above formalism to low temperature peaks 
in thermopower measurements under different ambient conditions 
it is important to account for different scattering mechanisms.  
Different scattering mechanisms in nanotubes can 
exhibit drastically different behavior, in analogy to the 
wide variety of phonon drag effects seen in conventional metals with 
different alloys.  In this work we have assumed that the 
electron relaxation time, $\tau$, is independent of temperature and 
that $\tau_p$ goes like $1/T^m$.
From Fig.~\ref{total_plot} we see that phonon-phonon 
scattering induces a peak in $S_{\text{total}}$.   
The $m=1$ case therefore appears be a good 
approximation for the samples of Ref.~\cite{romero}.  
Fig.~\ref{compare} also demonstrates that a temperature 
independent scattering mechanism $(m=0)$ produces only a 
knee in $S_{\text{total}}$.  Weak inflections in the thermopower
have also been observed in many 
thermopower measurements on carbon nanotubes.  
Other conditions may favor 
a decay mechanism which can significantly alter the 
temperature dependence of the drag thermopower.  
Detailed comparison with experiment 
will require suitable choices for the 
temperature dependence of the phonon lifetime.

We would like to thank H.E. Romero, P.C. Eklund, K. Park, 
P.E. Lammert and J.K. Jain for helpful 
discussions.  V. Scarola acknowledges support from the 
National Science Foundation grant no. DGE-9987589.

\begin{references}
\bibitem{caveat}As pointed out in Refs. \cite{cohen} and \cite{hone} the 
temperature dependence of the lattice specific heat depends on tube radius 
and inter-tube coupling.  Here, we consider a one dimensional model 
wherein interactions and finite tube radii do not induce a dimensional 
crossover in the lattice or electronic degrees of freedom.
\bibitem{caveat2} As long as the Fermi surface does not 
touch the zone boundary Umklapp scattering is 
exponentially suppressed at low temperatures \cite{blatt}.
\bibitem{limit} In the opposite limit, where 
electron-phonon scattering dominates phonon decay, one can show  
that at low temperatures, $S_{\text{drag}}$ is nearly 
independent of temperature.
\end{references}
\end{document}